\documentclass[aps,prl,twocolumn,superscriptaddress,amsmath,amssymb]{revtex4}

\usepackage{graphicx,amsmath,amssymb,wasysym,textcomp}
\usepackage{bm}
\usepackage{ulem}

\begin{document}
\title{Gauge fields and interferometry in folded graphene}
\author{Diego Rainis}
\email{diego.rainis@sns.it}
\affiliation{NEST, Istituto Nanoscienze-CNR and Scuola Normale Superiore, I-56126 Pisa, Italy}
\author{Fabio Taddei}
\affiliation{NEST, Istituto Nanoscienze-CNR and Scuola Normale Superiore, I-56126 Pisa, Italy}
\author{Marco Polini}
\affiliation{NEST, Istituto Nanoscienze-CNR and Scuola Normale Superiore, I-56126 Pisa, Italy}
\affiliation{Kavli Institute for Theoretical Physics China, CAS, Beijing 100190,  China}
\author{Gladys Le\'{o}n}
\affiliation{Instituto de Ciencia de Materiales de Madrid (CSIC), Sor Juana In\'es de la Cruz 3, E-28049 Madrid, Spain}
\author{Francisco Guinea}
\affiliation{Instituto de Ciencia de Materiales de Madrid (CSIC), Sor Juana In\'es de la Cruz 3, E-28049 Madrid, Spain}
\affiliation{Kavli Institute for Theoretical Physics China, CAS, Beijing 100190,  China}
\author{Vladimir I. Fal'ko}
\affiliation{Physics Department, Lancaster University, Lancaster LA1 4YB, UK}
\affiliation{Kavli Institute for Theoretical Physics China, CAS, Beijing 100190,  China}

\begin{abstract}
Folded graphene flakes are a natural byproduct of the micromechanical exfoliation process. In this Letter we show by a combination of analytical and numerical methods that such systems behave as intriguing interferometers due to the interplay between an externally applied magnetic field and the gauge field induced by the deformations in the region of the fold.
\end{abstract}

\maketitle

Gauge fields modulate the dynamics and interactions of electrons in a variety of scales from particle cosmology to
phase transitions in condensed matter~\cite{W08}. Graphene is a one-atom-thick carbon crystal where charge carriers behave
as massless quasiparticles~\cite{NGPNG09} and where the impact of lattice deformations on electrons is
equivalent to effective gauge fields~\cite{NGPNG09,GKG10,Betal10,VKG10}.
Graphene behaves as a highly stretchable membrane~\cite{membrane}, whose elastic deformations can be induced in a controlled way~\cite{deformations}. In this Letter we explore the most natural setup, a {\it graphene fold}~\cite{foldedgraphene} (see also~\cite{Getal07}), where the interplay between an applied magnetic field and the effective gauge fields leads to pronounced interference effects between chiral electronic modes. The device  allows for the controlled splitting of electronic current paths, opening the way for new multiterminal (Mach-Zehnder) interferometers~\cite{qhinterferometers}.

\begin{figure}[!h]
\begin{center}
\includegraphics[width=1.0\linewidth]{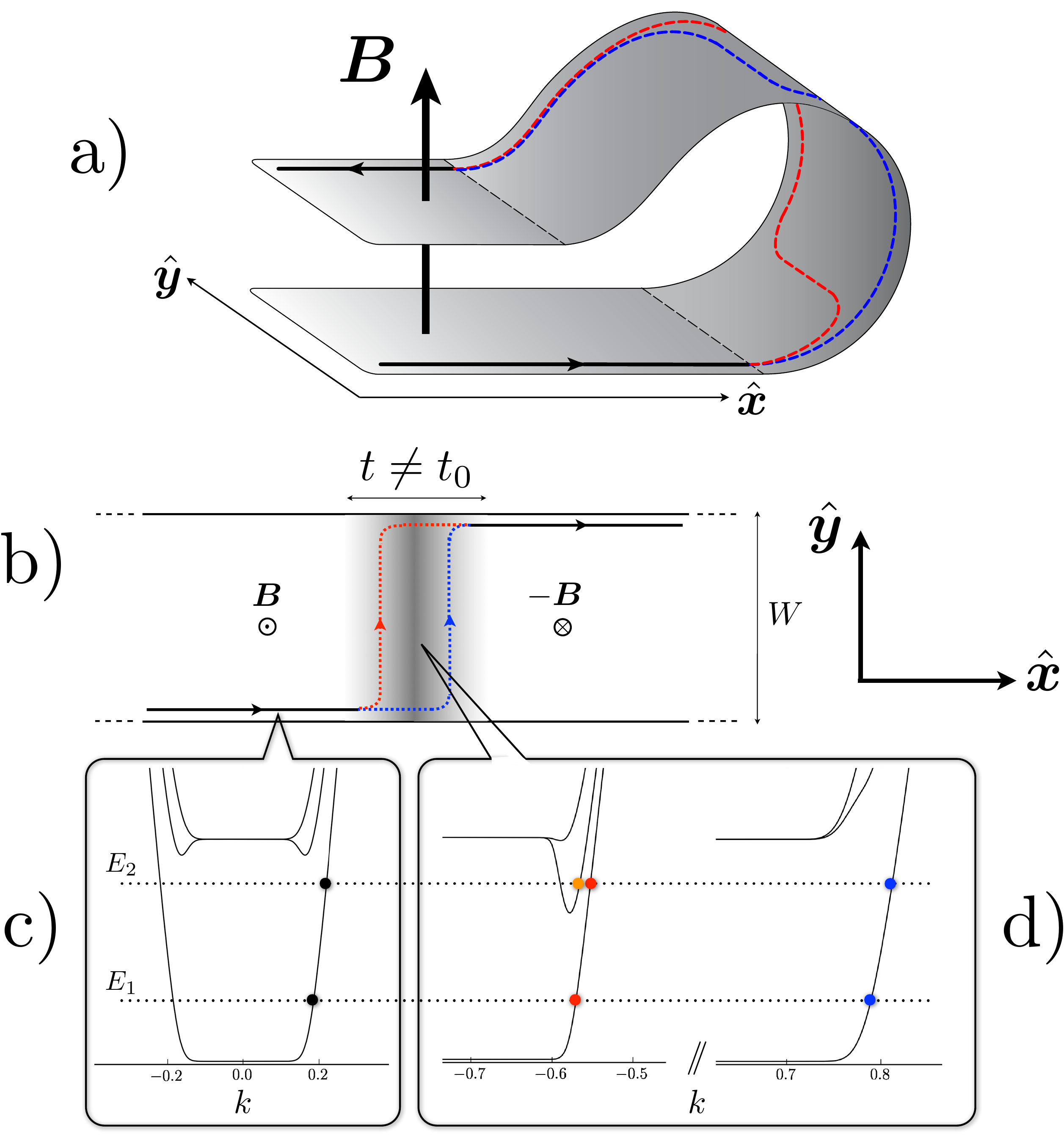}
\caption{\label{fig:one} (color online). a) A folded graphene ribbon in a magnetic field. b) A topologically equivalent geometry obtained by unfolding the ribbon. In this geometry it is clear that the magnetic field has opposite signs in the two layers. In the region of the fold (shaded area) two co-propagating snake states (red and blue lines), which are spatially-separated due to the deformation-induced gauge field, are present. c), d) The low-energy dispersion relations in the asymptotic leads [c)] and in the region of the fold [d)] are plotted as functions of Bloch momentum $k$ (in units of $\pi/a$, $a$ being the appropriate lattice constant).}
\end{center}
\end{figure}

We consider the setup sketched in Fig.~\ref{fig:one}a). It consists of a graphene nanoribbon with armchair edges (AGNR), which has been folded along a line perpendicular to its longitudinal axis, here taken to be along the ${\hat {\bm x}}$ direction. We assume that the two layers are completely decoupled~\cite{foldedgraphene} and that they are connected only through the region of the fold. Under these assumptions, current injected through a lead attached to the bottom layer, say, can be extracted from a lead attached to the top layer after having crossed the region of the fold.

The radius of the fold $R$ is determined by the balance between
the bending rigidity of graphene $\kappa$ and the van der Waals attraction $\gamma$ between the two layers~\cite{CSB09,ZXMMHZ10}. We find (see Supplementary Information) $R \approx \sqrt{\kappa/\gamma} \approx 7~{\rm \AA}$, where $\kappa \approx 1~{\rm eV}$ and $\gamma \approx 0.022~{\rm eV} \times {\rm \AA}^{-2}$. The deformation associated with the fold also induces a uniaxial strain along the ${\hat {\bm x}}$ axis, $u_{xx} \approx \gamma / ( \lambda + 2 \mu ) \equiv \bar{u} \approx 0.1\%$ where $\lambda \approx 3~{\rm eV} \times {\rm \AA}^{-2}$
and $\mu \approx 9~{\rm eV} \times {\rm \AA}^{-2}$ are the elastic Lam\'e coefficients of graphene.
This strain induces an inter-valley gauge field~\cite{SA02b,VKG10}, $A_x^{\rm str} = 0$,
$A_y^{\rm str} = \beta \bar{u} / a \approx 1.6 \times 10^{-3}~{\rm \AA}^{-1}$, where $a$ is the carbon-carbon distance ($a=a_0 \approx 1.42~{\rm\AA}$ at equilibrium), $\beta = - \partial \log(t) / \partial \log(a) \approx 2-3$, $t$ being the hopping between $\pi$ orbitals in nearest-neighbor carbon atoms ($t = t_0 \approx 3~{\rm eV}$ at equilibrium). The curvature of the fold hybridizes $\pi$ and $\sigma$ bands, leading to another contribution to the gauge field~\cite{KN08}, $A_x^{\rm hyb} = 0$, $A_y^{\rm hyb} = (3\epsilon_{\pi \pi} / 8\hbar v_{\rm F}) \times ( a / R )^2 \approx 7.2 \times 10^{-3}~{\rm \AA}^{-1}$, where $\hbar v_{\rm F} = 3 t a/ 2 \approx 6.4~{\rm eV} \times {\rm \AA}$ is the Fermi velocity multiplied by $\hbar$, $\epsilon_{\pi \pi} = V_{pp \pi} / 3 + V_{pp \sigma} / 2 \approx 3.0~{\rm eV}$, and $V_{pp \pi}, V_{pp \sigma}$ are hoppings between $p$ orbitals in nearest carbon atoms with different orientations. The smallness of the fold radius implies that the contribution from orbital hybridization is dominant. A gauge field of this magnitude varying over the length of the fold gives rise to an effective magnetic field with magnetic length $\ell_{B_{\rm eff}} \approx [(\pi R )/A_y^{\rm hyb}]^{1/2} \approx 5~{\rm nm}$, so that $B_{\rm eff} \approx 25~{\rm T}$.

Let us now consider the folded AGNR in  a perpendicular magnetic field ${\bm B} = B {\hat {\bm z}}$ [see Fig.~\ref{fig:one}a)], strong enough to quantize the single-particle spectrum into well-resolved~\cite{novoselov_science_2007} Landau levels (LLs), {\it i.e.} we assume that the magnetic length $\ell_B =\sqrt{\hbar c/(eB)}$ is much smaller than the width $W$ of the ribbon. Under these conditions, current in both layers is carried by edge states, which are localized on opposite sides of the sample for opposite current directions. Upon entering the region of the fold while moving along the transport (${\hat {\bm x}}$) direction, the out-of-plane component of the magnetic field first decreases, then becomes zero, and finally changes sign.

Since we are neglecting inter-layer hopping, our folded AGNR is topologically equivalent to an unfolded ribbon in the presence of a magnetic field step, as illustrated in Fig.~\ref{fig:one}b). Let us now imagine to inject current from left to right. In this case, current which is first carried by edge states near the bottom edge of the left layer will have to be carried, after crossing the fold, by edge states on the top edge of the right layer. At low energies (more precisely, at energies below the first LL plateau), only one edge state is involved in carrying this current. In the following we will restrict ourselves to this low-energy ``one-channel'' regime. Fig.~\ref{fig:one}c) shows a zoom of the dispersion relation $E = E(k_x)$ of an ideal AGNR in the presence of a quantizing magnetic field. Only the first few low-energy subbands forming  the zeroth and the first LLs have been plotted. Note that at sufficiently large values of $|k_x|$ these subbands become dispersive giving rise to the aforementioned quantum-Hall edge states.

Within (or in the proximity of) the region of the fold, a pair of transverse edge modes which live parallel to the axis of the fold (the ${\hat {\bm y}}$ axis) are induced by the change of sign of the magnetic field. These states are depicted by vertical blue and red lines in Fig.~\ref{fig:one}b). Their existence can be understood within a simple semiclassical picture. Similarly to what happens at the sample edges, skipping orbits form in each separate layer close to the ${\bm B}|-{\bm B}$ interface. These pair of skipping orbits living on the opposite sides of the interface merge together in a pair of so-called ``snake states''~\cite{snakestates}. In an ideal, flat, 
${\bm B}|-{\bm B}$ interface these two snake states are spatially superimposed and centered on the interface (which we define it to be the line where the ${\bm B}$ field vanishes).

The effect of the gauge field in the region of the fold is of paramount importance. It shifts the value of the momentum, and, as a result, it changes the position of the LL guiding center. The effective magnetic field has opposite sign for the two valleys, and is maximal in the armchair configuration, while it is zero in the zigzag case.

To understand better the nature of the snake states in the proximity of a ${\bm B}|-{\bm B}$ interface in the presence of deformations, 
we introduce an ``auxiliary system" consisting of a ${\bm B}|-{\bm B}$ interface infinitely extended along the ${\hat {\bm y}}$ direction, centered at $x=0$, and of finite length $w_{\rm tot}$ along the ${\hat {\bm x}}$ direction. We assume that a finite deformation-induced gauge field is present for $|x|< w/2 \ll w_{\rm tot}$. By looking at the spectrum $E(k_y)$ (Bloch momentum $k_y$ is a good quantum number because the auxiliary-system interface is infinitely extended along the ${\hat {\bm y}}$ direction) and at the eigenstates of this auxiliary system we can deduce some quantitative information about the snake states along the ${\hat {\bm y}}$ direction in our system, which has instead a finite transverse width.

Performing analytical calculations based on the continuum model and numerical ones based on a tight-binding Hamiltonian, we find the results plotted in Fig.~\ref{fig:one}d). Due to the deformation-induced gauge field the band structure $E(k_y)$ shows non-equivalent $K$ and $K'$ valleys. At low energies there is one dispersive edge state per valley [red and blue filled circles in Fig.~\ref{fig:one}d)]. 
Both edge states live close to the interface and are spatially separated by a distance $\delta_x$ given by
\begin{equation}\label{separation}
\delta_x \approx  A_y^{\rm hyb} \ell_B^2 \approx \frac{100~{\rm nm}}{B[{\rm T}]}~,
\end{equation}
using the parameters described earlier. [Eq.~(\ref{separation}) applies only for magnetic fields $B$ such that $\delta_x \lesssim \pi R$.] For realistic values of the magnetic field the separation between edge states is of the order of the length $\pi R$ of the folded region, and the carriers near the fold are valley polarized. This separation prevents short range scattering from mixing the valleys, unlike in other proposals to achieve valley polarization in graphene. The number of flux quanta in the region between the two channels, using the previous parameters, is
\begin{equation}\label{flux}
N_{\rm flux} \approx \frac{(\pi R) WB}{\Phi_0} \approx \frac{W [{\rm nm}] B[{\rm T}]}{1800}~,
\end{equation}
where $\Phi_0 \approx 4.1 \times 10^{3}~{\rm T} \times {\rm nm}^2$ is the magnetic flux quantum and we are neglecting a factor proportional to the angle between the field and the local orientation of the layer [we assume that the fold is not perfectly symmetric, see Fig.~\ref{fig:one}a)].

The flux in Eq.~(\ref{flux}) allows for the operation of the device in Fig.~\ref{fig:one}a) as an interferometer.
Incoming electrons propagating along the bottom edge state in the left layer, say, can either be transmitted to co-propagating edge states on the right layer or be reflected back. In both cases, the outgoing edge state is localized on the top edge of the sample. Electrons can move from one side to the other thanks to the existence of the snake states discussed above. Due to the gauge field-induced separation $\delta_x$ the two snake states  enclose a finite magnetic flux, Eq.~(\ref{flux}), which, in turn, can result in quantum interference.

We now turn to present our main numerical results for the differential conductance $G$ of a folded AGNR in a strong magnetic field and in the presence of deformations in the region of the fold. Our calculations, which rely on the Landauer formalism, are based on the tight-binding description and on the recursive Green's function technique~\cite{rainis_prb_2009}. 

\begin{figure}
\begin{center}
\includegraphics[width=1.0\linewidth]{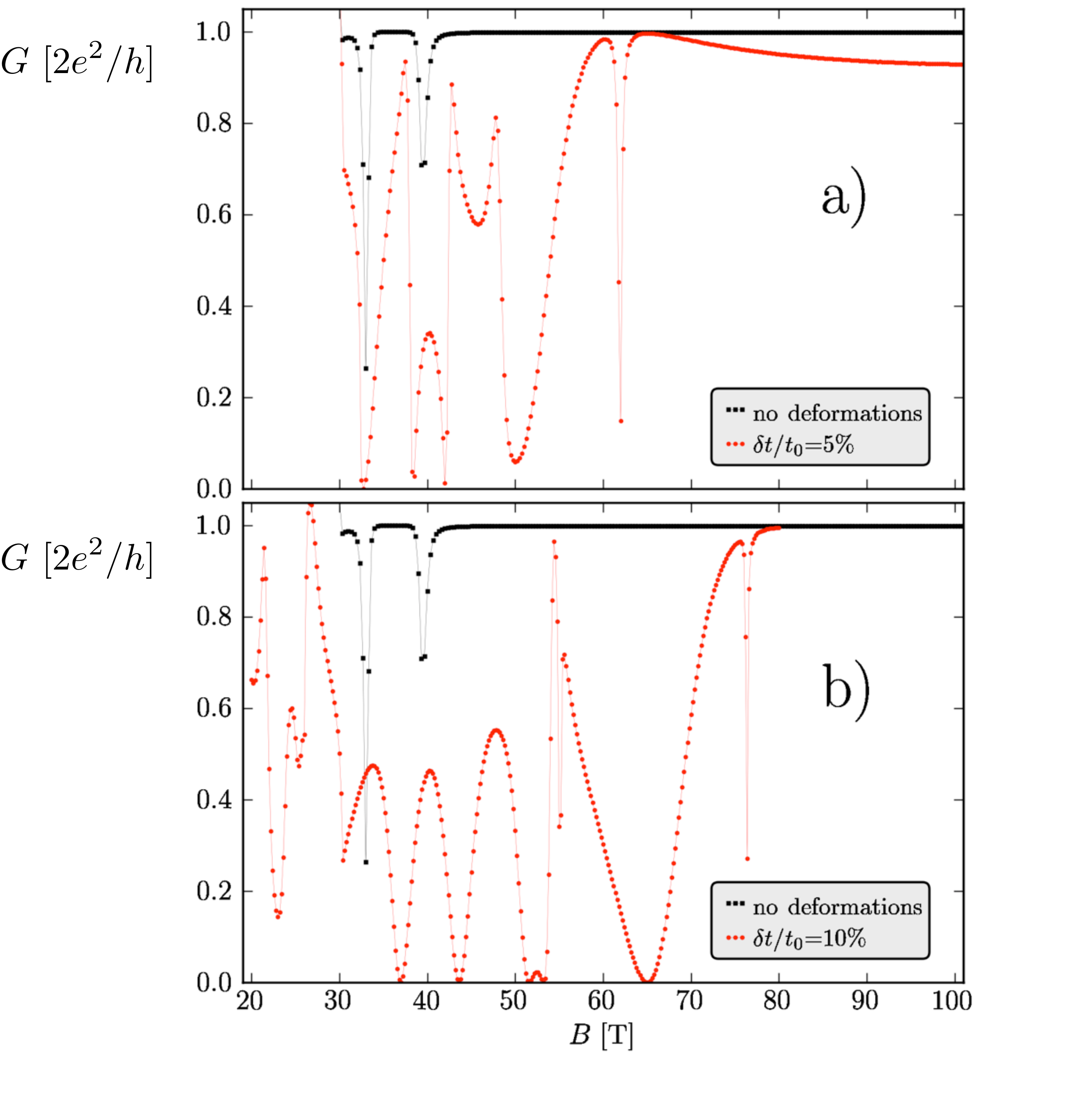}
\caption{\label{fig:two} (color online). (a) Numerical results for the two-terminal conductance $G$ (in units of $2e^2/h$) of the device in Fig.~\ref{fig:one}a) as a function of the applied magnetic field (in ${\rm T}$). In these simulations $W \sim 50~{\rm nm}$ and $w \sim 15$ nm. The chemical potential is located at $\approx 180~{\rm meV}$ above the Dirac level. The data labeled by filled (red) circles refer to the case of a folded AGNR with a modified hopping integral $t / t_0 = 0.95$ along the ${\hat {\bm x}}$ axis in the region of the fold. The data labeled by filled (black) squares refer to the case of a folded AGNR in the {\it absence} of deformations ($t / t_0 = 1.0$ everywhere). (b): Same as in a) but the data labeled by filled (red) circles have now been obtained for a value of $\delta t/t_0$ twice as big, {\it i.e.} $t / t_0 = 0.90$. }
\end{center}
\end{figure}

In Fig.~\ref{fig:two} we present data for $G$ as a function of the applied magnetic field $B$. The simulated ribbon has a width $W \sim 50~{\rm nm}$: for this size the quantum Hall regime is reached at $B\gtrsim 30~{\rm T}$. The window of magnetic fields displayed on the horizontal axis in this figure can be shrunk at will, provided that $W$ is increased (see Supplementary Information). The length $w$ of the folded region, in which the gauge field is active, is taken to be $w \sim 15~{\rm nm}$. We describe the total gauge field ${\bm A} = {\bm A}^{\rm str}+ {\bm A}^{\rm hyb}$ as a modulation of the hopping, although it is mostly due to the hybridization of the $\pi$ and $\sigma$ bands (see Supplementary Information). This description preserves the symmetries and main features of the gauge field, and should not change significantly the results. The modulation of the hoppings parallel to the ${\hat {\bm x}}$ axis, $\delta t / t_0 \equiv (t-t_0)/t_0$, varies between $5\%$ [Fig.~\ref{fig:two}a)] and $10\%$ [Fig.~\ref{fig:two}b)]. These values are  higher than those expected in a realistic fold. They allow us to illustrate the combined effect of gauge potentials and magnetic fields in ribbons of widths amenable to numerical analysis. A lower gauge field will give similar results in wider ribbons, see 
Eqs.~(\ref{separation}) and~(\ref{flux}).

The most important results of this work are the data represented by filled circles in Fig.~\ref{fig:two}. They clearly show an oscillatory behavior of the conductance $G$ as the magnetic field is varied. We emphasize that these oscillations are only due to the presence of a gauge field in the region of the fold. To make this more evident, in Fig.~\ref{fig:two} we have also reported the conductance of the same ${\bm B}|-{\bm B}$ interface {\it in the absence of deformations} (filled squares). We can clearly see how the conductance oscillations disappear when the gauge field is switched off. In this case, one can see that $G$ is, in the one-channel regime, always equal to the maximum conductance $2e^2/h$, apart from possible sharp resonances. This is part of a more general result: due to symmetry properties of the zeroth LL, a folded graphene nanoribbon {\it in the absence of a gauge field} shows always perfect transmission when current is carried by the zeroth edge state~\cite{prada_arXiv_2010}.

\begin{figure}
\begin{center}
\includegraphics[width=1.0\linewidth]{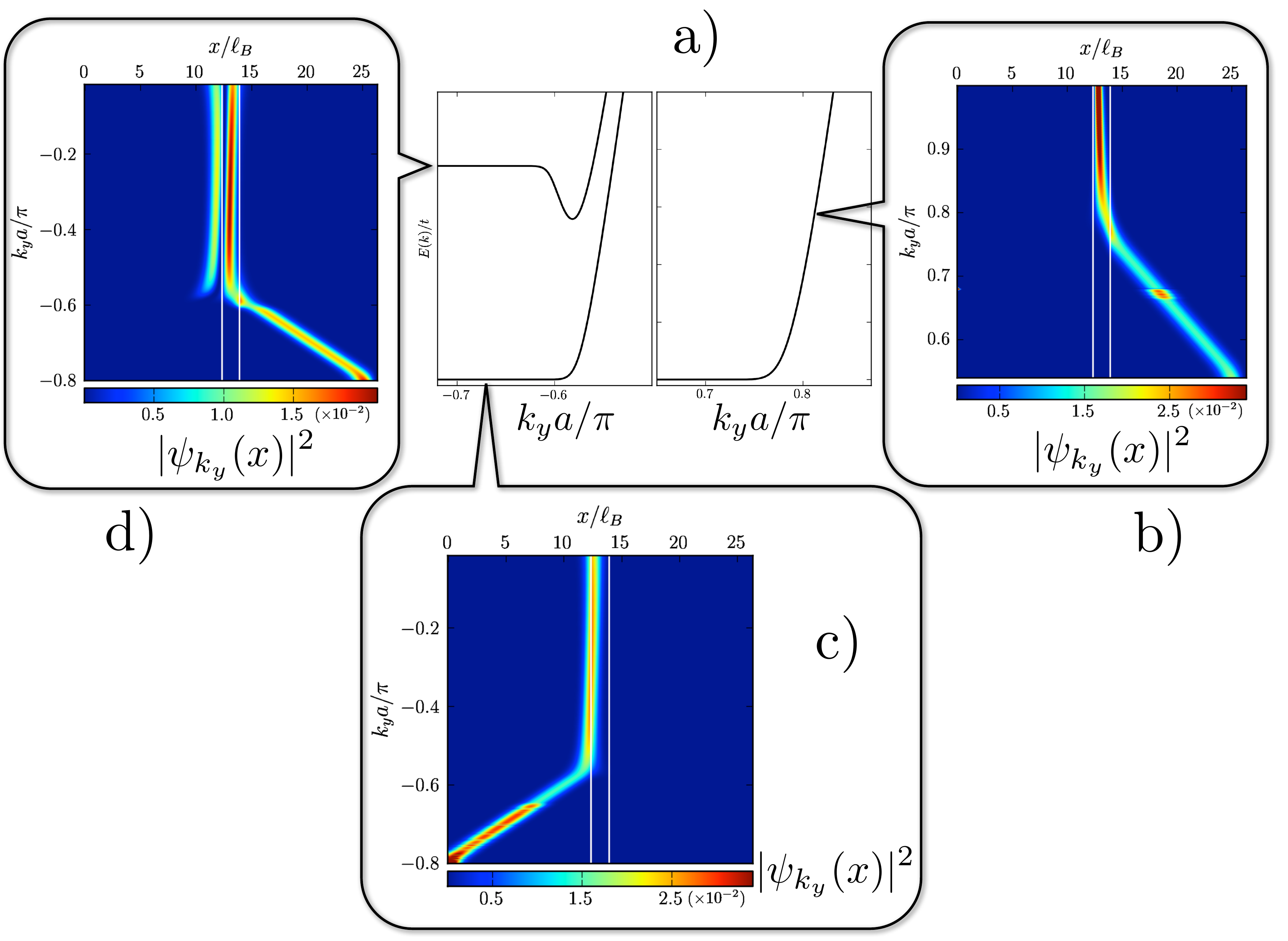}
\caption{\label{fig:three} (color online). a) A zoom of the dispersion relation of the auxiliary system (see main text) in the vicinity of the Dirac points. b) 2D color plot of the $K$-valley squared wave-functions $|\psi_{k_y}(x)|^2$ of the auxiliary system (only the value of the wave-function on one type of sublattice, say the $A$ type, is shown). In the horizontal axis the spatial position $x$ is shown in units of the magnetic length $\ell_B$. In the vertical axis we plot the Bloch momentum $k_y$ parallel to the interface. c) Same as in b), but for the $K'$ valley. d) Same as in the previous panels, but for the first-excited subband, which includes the first LL and the downward dip induced by the gauge field (only the $K$ valley is considered, the $K'$ valley has no dip).}
\end{center}
\end{figure}

Some comments related to the data in the presence of gauge fields are now in order.
i) Roughly, we can state that the behavior of the conductance as a function of the ${\bm B}$ field is characterized by two quite distinct regimes. We can identify a region of type ``I", approximately comprised between $30$ and $50~{\rm T}$, and a region of type ``II" for $B\gtrsim 50~{\rm T}$. (Note that for $B < 30~{\rm T}$ one exits the one-channel regime: for low fields the first LL energy is smaller than the chemical potential chosen in Fig.~\ref{fig:two}.) In region I the oscillations in $G$ are rather clear and regular, while in region II the conductance increases and then remains essentially large and constant for strong magnetic fields (modulo sharp Fano-like resonances). ii) Comparing the data in panel b) with those in panel a), we notice that a larger gauge field produces more regular and well-defined oscillations, but with smaller amplitude. Moreover, for larger gauge fields the type-II behavior is reached at larger values of the magnetic field.
Let us now examine more carefully the nature of the eigenstates of the auxiliary system introduced above. In Fig.~\ref{fig:three} we present numerical results for its eigenfunctions. Fig.~\ref{fig:three}a) shows a low-energy zoom of the dispersion relation $E(k_y)$ as a function of the Bloch momentum $k_y$ parallel to the ${\bm B}|-{\bm B}$ interface. We have selected only the Brillouin-zone portions related to the snake states we are interested in. We immediately observe that the existence of two adjacent regions with opposite magnetic fields modifies the usual LL structure of an ideal ribbon [Fig.~\ref{fig:one}c)], by introducing new dispersive branches which correspond to snake modes propagating along the interface. The presence of the gauge field has a crucial role in breaking the symmetry between the two valleys, as it generates a downward bending in the first LL subband of the $K$ valley, say, without modifying the same LL in the $K'$ valley.
Figs.~\ref{fig:three}b)-d) illustrate two-dimensional color plots of the envelope-function profile $\psi_{k_y}(x)$ as a function of the Bloch momentum $k_y$ and of the coordinate $x$ along the transport direction, for the three types of snake states we are considering. More precisely, Fig.~\ref{fig:three}b) shows the wave-function belonging to the $K'$ valley. For $0.55\apprle k_y a/\pi \apprle 0.75$ we have a standard zero-LL edge wave-function whose guiding center moves from the right side of the auxiliary system toward the center. For larger values of $k_y$ the wave-function becomes a snake state localized close to the interface. In Fig.~\ref{fig:three}c) we consider instead the $K$ valley, and we plot the wave-function profile corresponding to the lowest-energy subband. Again, for small values of $k_y$ we find ordinary edge states with a guiding center that moves from the left side of the auxiliary system towards the center. This is then converted into a snake state which approaches the interface asymptotically. It is crucial to observe that the ``centers" of the two snake states corresponding to the two valleys are {\it not} superimposed but are instead displaced by a finite amount. Finally, in Fig.~\ref{fig:three}d) we have plotted the wave-function $\psi_{k_y}(x)$ of the states belonging to the first LL. Since this is an excited subband, the corresponding wave-functions have a node, and thus the probability amplitude $|\psi_{k_y}(x)|^2$ presents a double-peaked structure in space, centered around the ${\bm B}|-{\bm B}$ interface. A more detailed analysis of the splitting of the LLs near the fold is given in the Supplementary Information.

The results presented here describe two terminal measurements in a folded graphene layer, and they assume that each layer can be contacted separately. The analysis can be extended in a straightforward way to multiterminal setups, where more complex correlations can be monitored. The  combination of flexibility and stiffness of graphene allows for many other combinations of gauge fields and electronic currents.

Work in Pisa and in Madrid was supported by the 2009/2010 CNR-CSIC scientific cooperation project. F.G. is supported by
MICINN (Spain), Grants FIS2008-00124 and CONSOLIDER CSD2007-00010.
F.T. acknowledges partial financial support by the Italian MIUR under the FIRB IDEAS project RBID08B3FM.
M.P., F.G., and V.F. also acknowledge partial financial support by the Project ``Knowledge Innovation Program" (PKIP)
of the Chinese Academy of Sciences, Grant No. KJCX2.YW.W10. D.R., F.T., and M.P. wish to thank Rosario Fazio for numerous enlightening discussions.

\begin{appendix}
\section{SUPPLEMENTARY INFORMATION}
This section contains technical details and extra numerical results relevant to the main text.
\section{Elastic strains and gauge fields}

The elastic energy of a graphene layer near a planar configuration can be written as~\cite{landau_book}
\begin{eqnarray} \label{elastic}
E_{\rm el} &=&\int d^2 {\bm r} \left\{ \frac{\lambda}{2} \left[ u_{xx} ( {\bm r} ) + u_{yy} ( {\bm r} ) \right]^2 +
\right. \nonumber \\
&+& \mu \left[ u_{xx}^2 ( {\bm r} ) + u_{yy}^2 ( {\bm r} ) + 2 u_{xy}^2 ( {\bm r} ) \right] \bigg\} + \nonumber \\
&+& \frac{\kappa}{2} \int d^2 {\bm r} \left[ \partial_{xx} u_z ( {\bm r} ) + \partial_{yy} u_z ( {\bm r} ) \right]^2~,
\end{eqnarray}
where $u_{ij}$ is the deformation tensor defined by~\cite{landau_book}
\begin{eqnarray}
u_{xx}  &=& \partial_x u_x + \frac{( \partial_x u_z )^2}{2} \nonumber \\
u_{yy}  &=& \partial_y u_y + \frac{( \partial_y u_z )^2}{2} \nonumber \\
u_{xy}  &=& \frac{\partial_x u_y + \partial_y u_x}{2} + \frac{\partial_x u_z \partial_y u_z}{2}~,
\end{eqnarray}
$u_i$ being atomic displacement vectors.
The parameters $\lambda \approx 3~{\rm eV} \times {\rm \AA}^{-2}$ and $\mu \approx 9~{\rm eV} \times {\rm \AA}^{-2}$ are the Lam\'e coefficients of graphene, 
while $\kappa \approx 1~{\rm eV}$ is the bending rigidity~\cite{ZKF09}. The van der Waals attraction energy between the graphene layers is
\begin{eqnarray}
E_{\rm vdW} = \gamma \int_\Omega d^2 {\bm r}~,
\end{eqnarray}
where $\Omega$ is the contact region, which does not include the fold, and $\gamma \approx 0.022~{\rm eV} \times {\rm \AA}^{-2}$ is the van der Waals interaction~\cite{SSG09}.

To a first approximation, the radius of the fold $R$ is determined by the balance between the bending rigidity and the van der Waals interaction~\cite{CSB09}. Approximating the fold by half a cylinder of radius $R$ we find
\begin{eqnarray}\label{erre}
R \approx \sqrt{\frac{\kappa}{\gamma}} \approx 7 ~{\rm \AA}~.
\end{eqnarray}
The presence of a substrate and the finite separation between the layers, $d \approx 3.3$~\AA, tend to increase this value and make the shape asymmetric. The distribution of curvatures and strains (see below) will become inhomogeneous and asymmetric as well.

Now, the bending energy is reduced by an expansion of the half cylinder, which leads to in-plane strains. Assuming a constant uniaxial strain, 
$u$, we find
\begin{eqnarray}
E_{\rm el} ( R ) = \frac{\pi R \kappa}{R^2 ( 1 + u )} + \frac{\pi R ( \lambda + 2 \mu ) u^2}{2}~.
\end{eqnarray}
Minimizing this expression with respect to $u$ and using Eq.~(\ref{erre}), we finally find
\begin{eqnarray}
u \approx \frac{\kappa}{( \lambda + 2 \mu ) R^2} \approx \frac{\gamma}{( \lambda + 2 \mu )} \equiv {\bar u} \approx 1.0 \times 10^{-3}~.
\end{eqnarray}
A similar result has been found for carbon nanotubes~\cite{SASRO99}. In the folded region, this strain leads to a constant gauge field~\cite{SA02b,M07,VKG10}
\begin{eqnarray}\label{strains}
\left \{
\begin{array}{ll}
A^{\rm str}_x = 0  \\
\\
\displaystyle {A^{\rm str}_y \equiv  \frac{\beta \bar{u}}{a}  \approx  1.6 \times 10^{-3} {\rm \AA}^{-1} }
\end{array}\right.~,
\end{eqnarray}
where $a$~is the distance between nearest neighbors in the graphene lattice ($a=a_0 \approx 1.42~{\rm\AA}$ at equilibrium), $\beta = - \partial \log ( t ) / \partial \log (a) \approx 2 -
3$, and $t$ is the hopping between $\pi$ orbitals in nearest neighbor atoms ($t = t_0 \approx 3~{\rm eV}$ at equilibrium).
This situation has been considered in Ref.~\onlinecite{FGK08}. In the absence of a magnetic field, the transmission can be calculated analytically.

The bending of the layer induces the hybridization of $\pi$ and $\sigma$ orbitals, and the emergence of a new contribution to the total gauge field, not related to strains~\cite{KN08}. 
Its value is
\begin{eqnarray}\label{hybrid}
\left \{
\begin{array}{ll}
A^{\rm hyb}_x = 0  \\
\\
\displaystyle {A^{\rm hyb}_y \equiv  \frac{3 \epsilon_{\pi \pi}}{8 \hbar v_{\rm F}} \frac{a^2}{R^2} \approx  7.2 \times 10^{-3} {\rm \AA}^{-1} }
\end{array}\right.~,
\end{eqnarray}
where $\hbar v_{\rm F} = 3 t a / 2 \approx 6.4~{\rm eV} \times {\rm \AA}$ is the Fermi velocity multiplied by $\hbar$ and $\epsilon_{\pi \pi} = V_{pp \pi}/3+V_{pp \sigma}/2 \approx 3.0~{\rm eV}$, $V_{pp \pi}$ and $V_{pp \sigma}$ being the two possible hopping integrals between $p$ orbitals in neighboring carbon atoms. Due to the smallness of the radius of the fold, the contribution from the hybridization in Eq.~(\ref{hybrid}) is roughly one order of magnitude larger than the one due to strains, given in Eq.~(\ref{strains}). It is equivalent to a relative change of the value of the hopping $t$ of $1\%$.
\begin{figure}[h]
\begin{center}
\includegraphics[width=1.0\linewidth]{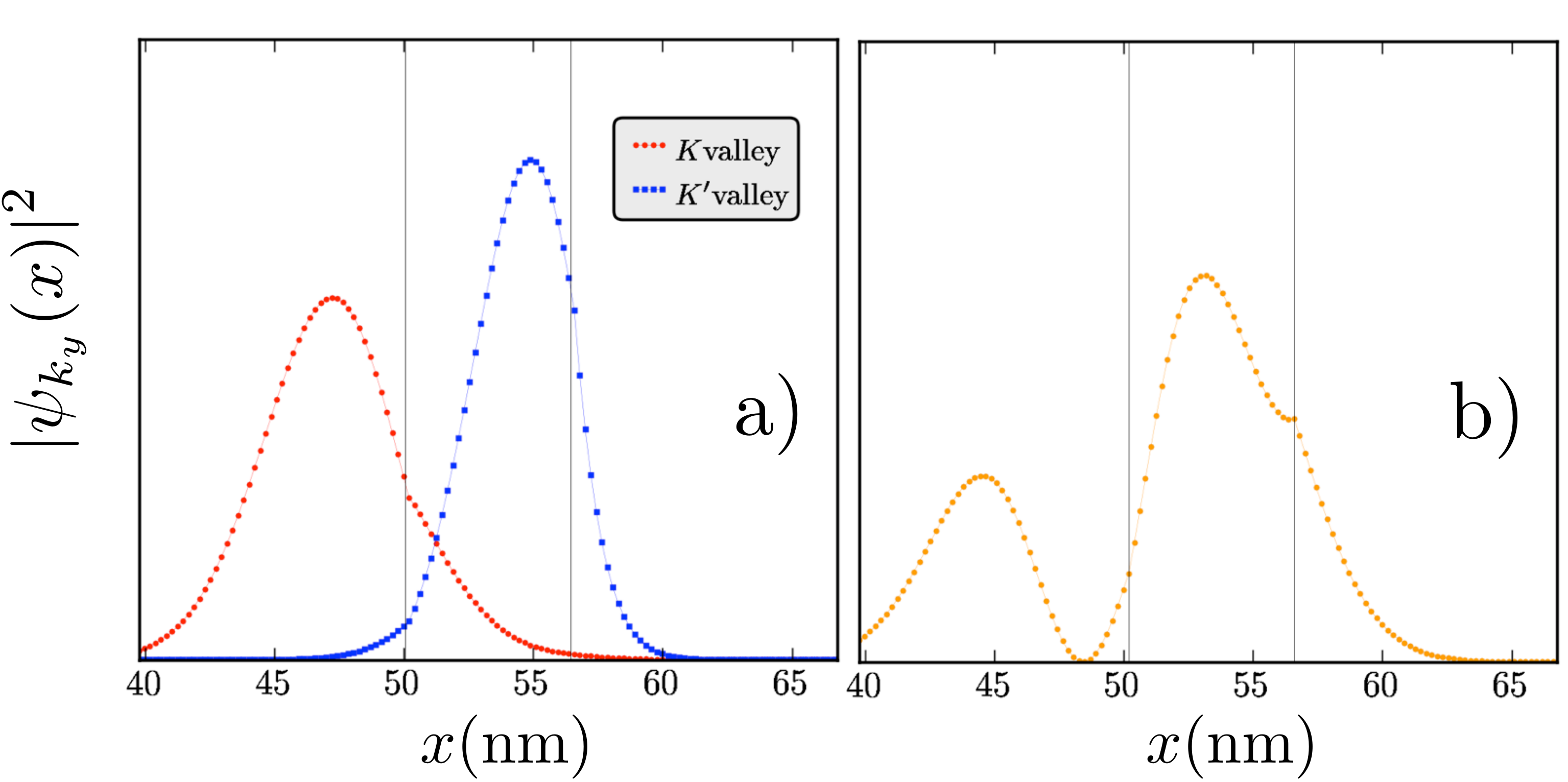}
\caption{Wave-functions in the ${\bm B}|-{\bm B}$ interface in the presence of deformations. Panel a) Probability density $|\psi_{k_y}(x)|^2$ for the snake states residing along the ${\bm B}|-{\bm B}$ interface. The horizontal axis is a zoom around the center of the auxiliary system where the interface is located. Data labeled by filled red circles refer to the $K$ valley, while data labeled by filled blue squares refers to the $K'$ valley. The two vertical thin lines define the deformed region within which the ${\bm B} \to -{\bm B}$ transition takes place. We have chosen an intermediate energy value in the one-channel regime, while the magnetic field is $B = 40~{\rm Tesla}$.  Panel b) Probability density $|\psi_{k_y}(x)|^2$ for the snake state belonging to the first-excited subband [see Fig.~3d) in the main text]. Notice that this single eigenstate is characterized by a double-peak probability distribution, which could generate interference effects similar to the ones occurring
in an Aharonov-Bohm ring inteferometer.\label{fig:onesupplementary}}
\end{center}
\end{figure}

\section{Snake states}
\begin{figure}[h!]
\begin{center}
\includegraphics[width=1.0\linewidth]{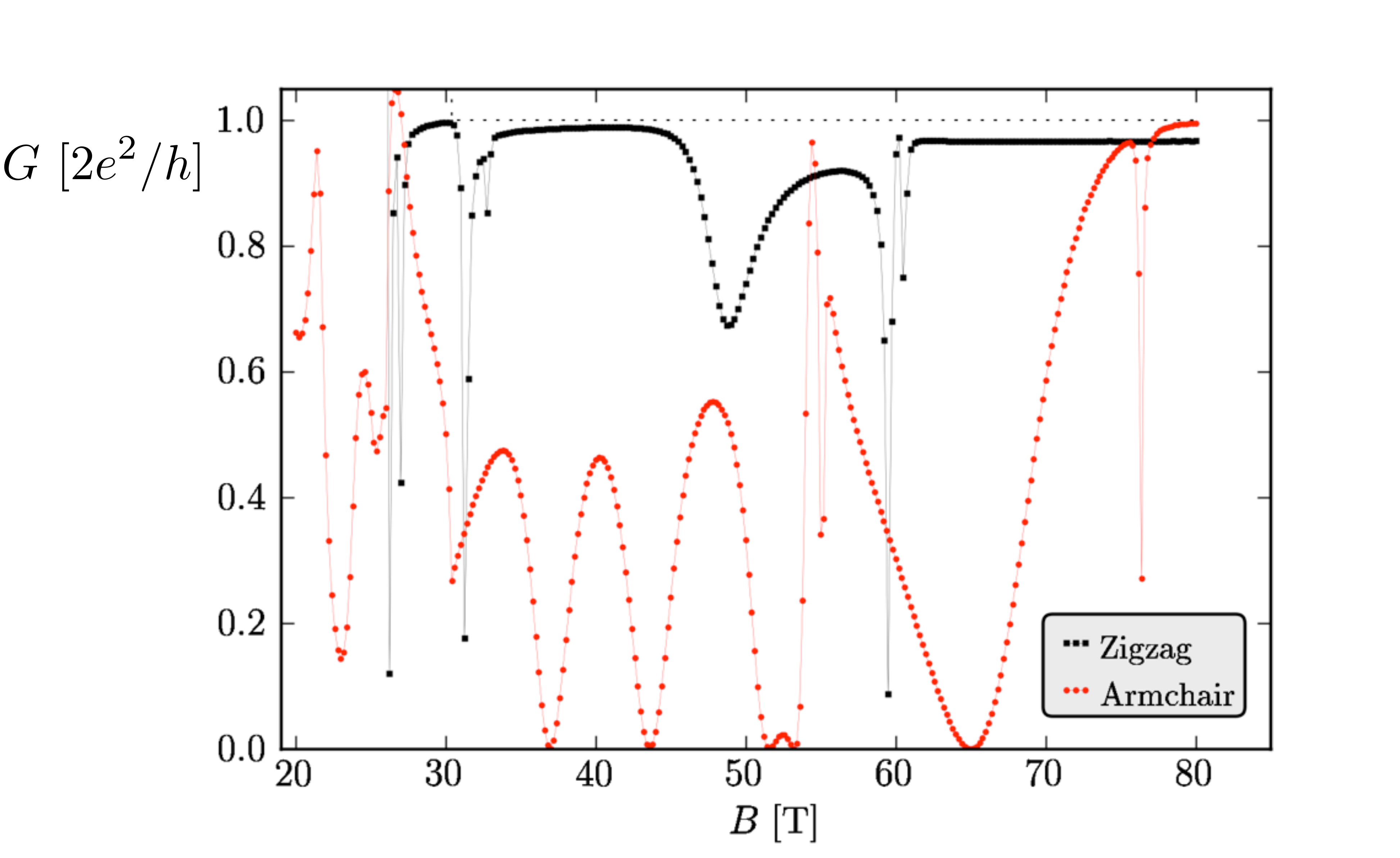}
\caption{Conductance through a nanoribbon with zigzag edges. Plot of the two-terminal conductance (in units of $2e^2/h$) 
of the folded graphene nanoribbon with zigzag edges along the transport ($\hat{\bm x}$) direction. 
The parameters used to produce this plot are identical to those used in Fig. 2 of the main text. For the sake of comparison, the data for armchair edges have been reported again. It is evident that in the case of zigzag edges no regular and distinct oscillations are present, the conductance being almost always equal to unity apart from some resonances and minor variations.\label{fig:twosupplementary}}
\end{center}
\end{figure}
Some of the important features that might be hidden in the color plots of Fig.~3 in the main text have been highlighted here in Fig.~\ref{fig:onesupplementary}, which
contains typical one-dimensional cuts of $\psi_{k_y}(x)$ for a fixed value of $k_y$. More precisely, Fig.~\ref{fig:onesupplementary}a) reports the squared modulus $|\psi_{k_y}(x)|^2$ of the $A$-sublattice wave-function corresponding to a snake state in the lowest-energy subband. The spatial region where deformations are present (here taken to be $6~{\rm nm}$ wide) is inside the area delimited by the two vertical lines. The position where the ${\bm B}$ field vanishes in not shown in this plot for simplicity but lies within this region. The value of $k_y$ chosen to make this plot corresponds to an energy intermediate between the zero and the first LL. As shown in Fig.~3 of the main text, one of these two snake states belongs to the $K$ valley, while the other one belongs to the $K'$ valley. From Fig.~\ref{fig:onesupplementary}a) it is very clear that the centers of the two snake states are slightly shifted with respect to each other by roughly $10~{\rm nm}$. Some finite magnetic flux can thus be enclosed between them, assuming that their spatial displacement is not perfectly symmetric with respect to the position where ${\bm B}$ vanishes (this is necessary to avoid a perfect -- and unrealistic -- cancellation between the fluxes coming from the ${\bm B}$ and $- {\bm B}$ regions). This magnetic flux is at the origin of interference effects and conductance oscillations shown in Fig.~2 of the main text. If the deformed ${\bm B} | -{\bm B}$ interface provides a finite coupling between the two 
snake states belonging to opposite valleys, the enclosed residual flux manifests itself as a phase difference between the two quantum trajectories involving the $K$ and $K'$ interface snake states. We believe that this is the mechanism responsible for the conductance oscillations of type II shown in Fig.~2 of the main text.

Another appealing possibility of interference is offered by the snake state depicted in Fig.~\ref{fig:onesupplementary}b), which belongs to the first subband of the $K$ valley. As we have already pointed out, this state presents a node close to the interface, and thus the corresponding probability density $|\psi_{k_y}(x)|^2$ possesses a double-peak structure. In this case thus there is no need of an inter-valley coupling mechanism to observe interference. An incoming electron can split at the interface into a two-path configuration, which can generate Aharonov-Bohm-type  interference analogous to the one seen in ordinary quantum rings. We have checked that the type-I oscillations in Fig. 2 of the main text occur precisely in the magnetic-field range where this excited snake state exists at the chosen chemical potential.

\section{Influence of lattice orientation}

As already shown in Fig.~2 of the main text (data labeled by black filled squares), the conductance oscillations discussed above are absent when the gauge field is switched off. Our results then are similar to those reported in Ref.~\onlinecite{prada_arXiv_2010}. 

A similar disappearance of the oscillations occurs if the orientation of the lattice with respect to the fold axis is rotated by 90$^\circ$ (zigzag edges along the transport direction). 
The gauge field in this case is ${\bm A} =  A(x) {\hat {\bm x}} \propto \Theta(w/2-|x|) {\hat {\bm x}}$. Such a pseudo-vector potential has zero curl, {\it i.e.} it can be gauged away. We thus expect that the oscillations disappear in this case: this is indeed confirmed by our numerical calculations, as illustrated in Fig.~\ref{fig:twosupplementary}. It is interesting to note that the gauge transformation is applicable only to the effect of the gauge field on the electronic wavefunctions. The lattice deformations which originate the gauge field, obviously, cannot be cancelled.

\section{Numerical results for wider ribbons}
\begin{figure}[h!]
\begin{center}
\includegraphics[width=1.0\linewidth]{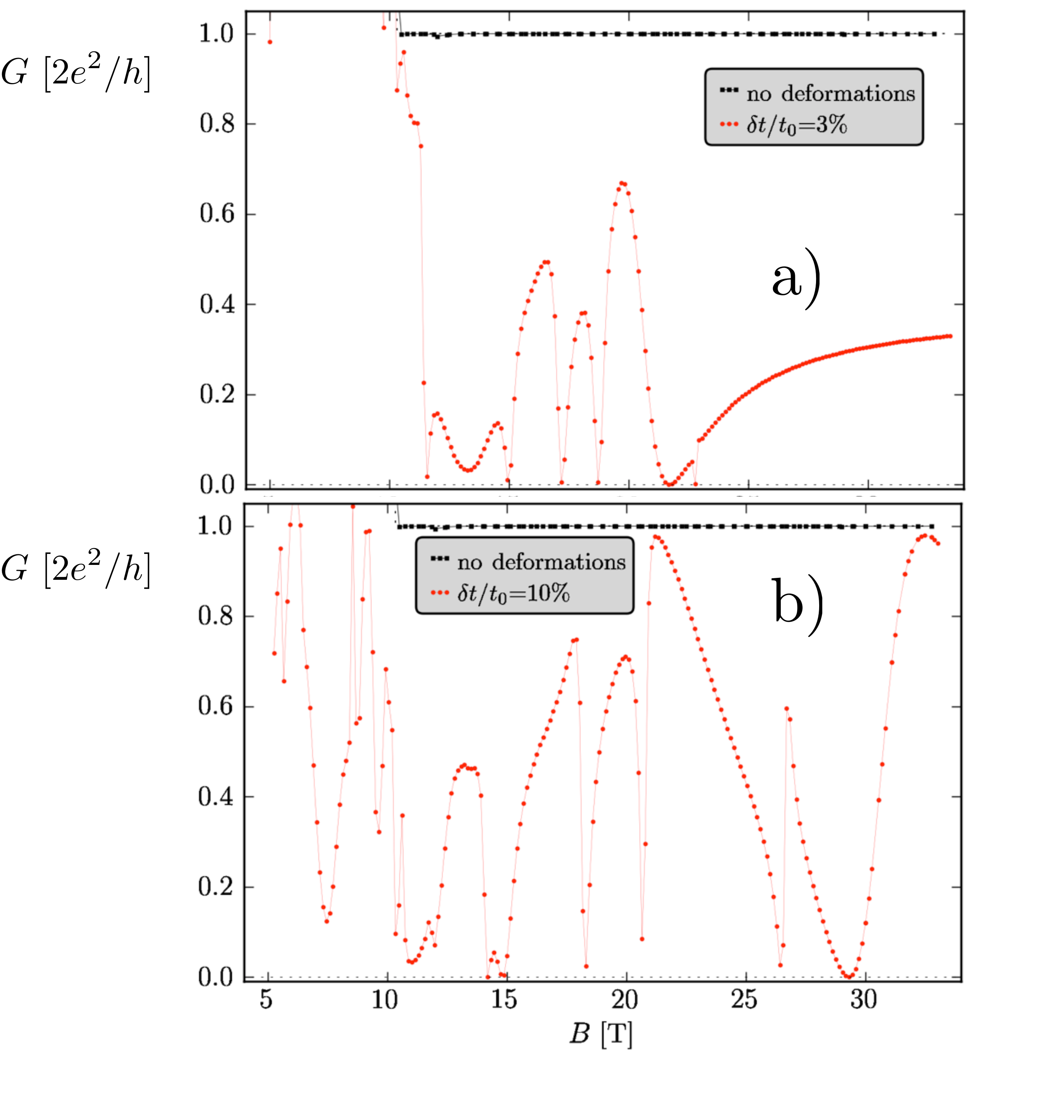}
\caption{Gauge-field-induced conductance oscillations through a wider ribbon. Plot of the two-terminal conductance of an armchain-folded-graphene nanoribbon with $W = 100~{\rm nm}$. 
The region where deformations are present is $30~{\rm nm}$ long, the chemical potential has been set to $\approx 100~{\rm meV}$ above the Dirac level, and the hopping variation is $\delta t /t_0 = 3\%$ [panel a)] and $10\%$ [panel b)]. Color and symbol coding are identical to those used in Fig.~2 of the main text.\label{fig:threesupplementary}}
\end{center}
\end{figure}
As already mentioned in the main text, most of our numerical calculations have been carried out in graphene ribbons with width $W=50~{\rm nm}$ solely for computational reasons. 
The corresponding quantizing magnetic field, given by the requirement $\ell_B \ll W$, turned out to be $B\gtrsim 30~{\rm Tesla}$.  
The range of magnetic fields where oscillations in the conductance are seen (see Fig.~2 in the main text) 
is largely out of experimental reach. Nevertheless, the same analysis can be carried out for wider ribbons. Here we have chosen a ribbon 
with a transverse width of $W=100~{\rm nm}$ for which the quantum Hall regime is reached at $10~{\rm Tesla}$, 
the conductance oscillations occuring in a range of magnetic field values ($10 < B < 30~{\rm Tesla}$) that is currently experimentally accessible. 
In Fig.~\ref{fig:threesupplementary} we report conductance data for this system. The value of $\delta t/t_0$ is $3\%$ in panel a) and $10\%$ in panel b), 
while the length $w$ of the folded region is taken to be $w = 30~{\rm nm}$. 
We clearly see that Fig.~\ref{fig:threesupplementary} shows a pattern of oscillations which shares the same basic physical features with the data reported in Fig.~2 of the main text.

\end{appendix}

\end{document}